# Network Load Balancing Methods: Experimental Comparisons and Improvement

Shafinaz Islam
Electrical and Electronic Engineering
Rajshahi University of Engineering and Technology Rajshahi, Bangladesh
shafinaz.orthy16@gmail.com

*Abstract*—Load balancing algorithms play critical roles in systems where the workload has to be distributed across multiple resources, such as cores in multiprocessor system, computers in distributed computing, and network links. In this paper, we study and evaluate four load balancing methods: random, round robin, shortest queue, and shortest queue with stale load information. We build a simulation model and compare mean delay of the systems for the load balancing methods. We also provide a method to improve shortest queue with stale load information load balancing. A performance analysis for the improvement is also presented in this paper.

Keywords: *Load balancing, Load balancing simulation, Radom, Round robin, Shortest queue.*

## I. INTRODUCTION

Load balancing is a significant component of current network infrastructure and computer systems where resources are distributed over a large number of systems and have to be shared by a large number of end users. Load balancing is the way of distributing workload across resources in order to get optimal resource utilization, minimum response time or reduced overload.

Load balancing is also a basic problem in many practical systems in daily life. A common example may be a supermarket model where a central balancer or dispatcher assigns each arriving customer to one of a collection of servers to minimize response time.

It is the task of the load balancers to determine how jobs will be distributed to resources. To do this, load balancers use some type of metric or scheduling algorithm. Load balancers may use the state of the server for scheduling strategy or may not. If the load balancer does not use state of the resource then it is called stateless balancing strategy. One example of stateless load balancing strategy may be random load balancing where load balancer distributes the workload across servers randomly, meaning it picks a server randomly. Round robin load balancing is also a stateless load balancing strategy where workloads are distributed in round robin fashion. On the other hand, if load balancer uses information of resource then the balancing strategy is called stateful. One type of stateful load balancing is shortest queue. The load balancer gets the updated queue length of resources and assigns a job to a resource that has shortest queue length.

In this paper we evaluate four types of load balancing methods: random, round robin, shortest queue and shortest queue with stale load information. Also, we explore how load balancing given out-of-date state information can be improved over a simple shortest queue method. The primary performance measure of interest is mean delay for an arriving job to have completed service. Another measure of interest is the utilization of resources. To do this, we build a simulation model using CSIM20 [1]- a development toolkit for simulation and modeling. We run experiments and present the result in this paper.

## II. RELATED WORKS

In [2] Zhou presents trace driven simulation study of dynamic load balancing in homogenous distributed system. From production system they have collected job's CPU and I/O demand and used them as input of the simulation model, which includes representative CPU scheduling policy. They have simulated seven load-balancing algorithms and compared their performances. They conclude that periodic and non-periodic information exchange based algorithms provide similar performance and the periodic policy based algorithms that use salient agent to collect and distribute load information decrease overhead and provide better scalability. Eager et al [3] explores system load information in adaptive load for locally distributed system. They studied three simple abstract load sharing policies (Random, Threshold, Shortest), compared their performance to each other and to two bounding cases such as no load sharing and perfect load sharing. The result suggests that simple load sharing policies using small amount of state information performs dramatically better than no load sharing and nearly as well as the policies that use more complex information. Dahlin examines [4] the problem of load balancing in large systems when information about server load is stale. The paper develops strategies that interpret load information based on its age. The crux of the strategy is to use not only the stale load information of the server, but also to use the age of the information and predict the rate at which new information arrives to change that information. Load balancing in dynamic distributed system in cases of old information, has been also studied by Mitzenmacher [5]. The paper concludes that choosing the least loaded of two random choices performs better over large range of system parameters and better than similar strategies, in terms of expected response time of the system.

Gupta et al [6] presents an analysis of join-the-shortest-queue routing for web server farms. A server farm consists of a router, or dispatcher, which takes requests and distributes jobs to one of collection of servers for processing. They presented the idea of Single-Queue Approximation (SQA) where one designated queue in the farm is analyzed in isolation using state dependent arrival rate. They conclude that SQA, in some sense, capture the effect of other queues.

For a system of two or more servers where jobs arrive according to some stochastic process and if the queue length is the only system information known, then dispatching the job to the server with the shortest queue length is one of the ways to minimize the long run average delay to service each customer. But in [7] Whitt shows that there are service time distributions for which it is not optimal to dispatch the job to the server that has shortest queue length. Also, if the elapsed service time of customers in service is known, in the long run dispatching job to the server that minimizes their individual expected delay does not always optimize average delay time.

Soklic [8] introduces diffusive load balancing algorithm and compares its performance with three other load balancing algorithms such as static, round robin, and shortest queue load balancing. The result shows that diffusive load balancing is better than round robin and static load balancing in a dynamic environment.

III. MODEL AND NOTATION

Our system consists of a front-end dispatcher (load balancer) that receives all incoming jobs, and five similar servers. Each server has its own queue, where the job is put if the server is currently busy. Jobs arrive to the dispatcher and are immediately transferred to one of the servers where they join the server's queue.

The following assumptions for the system are made:

- Server performs a job at first-come-first-serve basis.
- Jobs arrive according to a Poisson process with the rate equals to $\lambda$.
- Service times are independent and exponentially distributed with a mean 1.0 second.
- Each server has the same service rate equals to $\mu$.
- When a job comes, the dispatcher is able to make a decision immediately at wire speed. That means that it takes zero time to the dispatcher to make a decision to which server to transfer a job.
- A job can be transferred to a server in zero time with zero cost.
- Server's queue is not infinite. When there are more than 200 jobs in any queue, the system is considered to be broken.
- When the job is sent to the server, it cannot resent in to another server. It means that only the dispatcher can make a decision to which server a job will be sent.
- Jobs are not preemptible. It means that we cannot interrupt a task if the server starts to process it.

The dispatcher implements some load balancing algorithms, which chooses the server to transfer the job. According to [2], a load balancing algorithm in general consists of the three components:

*1)* The *information policy* specifies what information is available to the dispatcher when it makes the decision, and the way by which this information is distributed.
*2)* The *transfer policy* determines the eligibility of a job for load balancing.
*3)* The *placement policy* chooses for the eligible job a server to transfer this job to.

We assume that every job is eligible for load balancing and, thus, will be transferred to the server by the dispatcher. The information policy and the placement policy will be defined in the Section IV for each of the load balancing strategies.

IV. LOAD BALANCING STRATEGIES

In this paper we will consider two main types of the load balancing strategies that uses information about the system to make a decision:

- With *up-to-date state information*. That means that the dispatcher has relevant information every time it makes a decision.
- With *stale information*. That means that information is updated periodically and the dispatcher won't operate with the recent information each time it has to transfer a job.

The case with the stale information is more realistic because in the real system we have different latencies for each server and it is extremely hard to get all the information quickly and in one time. Also, requesting each server every time the job arrives generates a lot of additional traffic in the network. Thus, strategies with the stale information scale better which is very important for the real system.

*A. Random strategy*

In the Random load balancing strategy an incoming job is sent to the Server $i$ with probability $1/N$, where $N$ is the number of servers. This strategy equalizes the expected number of jobs at each server. Within Random load balancing strategy the dispatcher has no information about the job or the server's states.

*B. Round Robin strategy*

In the Round Robin strategy jobs are sent to the servers in a cyclical fashion. That means that $i$-*th* job is sent to the Server $i \bmod N$, where $N$ is the number of servers in the system. This strategy also equalizes the expected number of jobs at each server. According to [9], it has less variability in the job's interarrival time for each server than the Random strategy. Within Round Robin load balancing strategy the dispatcher stores the ID of the server to which the last job was sent.

*C. Shortest Queue strategy with up-to-date information*

In the Shortest Queue strategy the dispatcher sends job to the server with the least number of jobs in the queue. If there is several servers with the smallest queue length, then the dispatcher randomly choose a server form this list. This strategy tries to equalize the current number of jobs at each server.

Within this Shortest Queue strategy the dispatcher gets the length of the servers queue every time it makes a decision. Further, the length of the queue is considered to be the number of jobs in the server and the server's queue.

Winston [10] shows that the Shortest Queue strategy is optimal if the following conditions are met:

- There is no prior knowledge about the job.
- Jobs are not preemptible.
- Dispatcher makes a decision and transfers a job immediately when it comes.
- Each server processes task in First-Come-First-Served order.
- The job size has exponential distribution.

In his work, Winston defines the optimality as maximizing the number of jobs completed by some constant time T. So, as our model meets all the requirements, the Shortest Queue with the up-to-date information can be considered as the optimal strategy. The problem is that we cannot use this strategy in the real system because we cannot provide the information about the system state each time the dispatcher makes a decision.

Later in the paper we will sometimes refer to this strategy as an USQ strategy.

*D. Shortest Queue strategy with stale information*

In this Shortest Queue strategy, the dispatcher also sends job to the server with the least number of jobs in the queue. The difference from the strategy C is that the information about the system's state is updated periodically. So, the dispatcher doesn't have up-to-date information every time it sends the job to the server. Ties within this strategy are broken in the same way: if there are several servers with the smallest queue length, then the dispatcher randomly choose a server form this list.

As we said, within this Shortest Queue strategy the dispatcher also uses the lengths of the server queues when it makes the decision. But in this case information is not recent and is updated on periodical basis.

Later in the paper we will sometimes refer to this strategy as an SSQ strategy.

*E. Shortest Queue strategy with stale information and keeping the history of the jobs scheduling during the stale period*

As we will see in the Section VI, the shortest queue strategy works much worse when it operates with the stale information comparing to its results with up-to-date information. Therefore, we come up with the new strategy that helps us to improve the load balancing result when information is updated periodically.

The main idea of this strategy is to keep history of the jobs scheduling during the stale period. It means that during the stale period we keep track to which server we sent a job and our next decision is based on not only on the queue lengths information but also on this history. More precisely, we increase the known queue length for the server each time we sent a job to it. When the information update comes, it rewrites the length of the queues and thus, deletes our history for that stale period. After the update we start our history from scratch.

To keep the history in this way, we even don't need addition memory because we use the array where we keep queues length. So, the history doesn't take a lot of memory and, as we will see, can significantly increase the performance of the dispatcher.

The pseudo-code of the algorithm that implements this strategy is listed below:

*GENERATE random number for each server;*
*FIND the server with the shortest queue;*
*IF several servers have the shortest queue length)*
*THEN pick one with the least generated random number);*
*INCREASE the length of the queue for the chosen server by 1;*
*RETURN the ID of the chosen server as a result;*

As we mentioned above, an update overwrites the information about the queue lengths every time.

Within this strategy the dispatcher also operates only with the servers queue length information that is updated periodically. Later in the paper we will sometimes refer to this strategy as an HSQ strategy.

## V. EXPERIMENTS DESIGN

There are a lot of tools that can help to build the simulation model. In this study we use CSIM 20 for C to build the model. CSIM is a function library that provides the developer with numerous methods and data structures to simplify the implementation of the model. The first version of the library was announced in the 1980s. So, it is very stable. Also, as a C library, it provides with great flexibility during the project development.

During the experiments performance of the model is measured by the mean response time for the job. The job's response time is defined as the simulation time between the job's arrival and completion at one of the servers. We want to mention that in the experiments CSIM gives us an opportunity to work with the simulation time instead of the CPU time. This makes our measurement more precise and independent from the machine we use.

In the experiments we want to study the influence of the following factors:

- *Load balancing strategy*. Within this project we compare five different strategies listed in the Section IV.
- *System workload*. System workload is defined as the ration of the arrival rate $\lambda$ to the service rate $5\mu$ and is usually represented using percentage. It is easy to understand that if the arrival rate is greater than the service rate, then system is not stable. The number of the jobs in the queue will grow infinitely and the system will be broken after some queue will hit the 200 jobs in the queue threshold. Within this project we will change the workload from 0% to 90%. We won't consider a workload more than 90% because it cannot be met in the real system as someone will likely increase the number of servers before the workload will be this high.
- *Length of the stale period*. Length of the stale period is the time between system's information update. In

this project we will change this factor from 1 second to 25 seconds to study its influence.

We will use factorial design of the experiments. Within this project we want to do the following:

- Compare the performance of different load balancing strategies.
- Understand how the performance of the system depends on the system workload.
- Understand how the length of the stale period decreases the performance of the load balancing strategy.
- Examine does keeping history in the shortest queue strategy help us to decrease the mean response time of the system?

To make the results significant we will use the CSIM approach that is called run length control. Using it, we will stop the simulation when the results archive the desired accuracy with 95% probability. In this project the error of 0.01 during measurements can be seen as insignificant so, this accuracy will be used.

As we said, the results do not depend on the machine used. But, to be precise, we used a machine with Intel i5 processor, 4GB RAM with Windows 7 Home Edition on it. To build the model we used C in Visual Studio 2010 Professional Edition and the CSIM 20 for C library.

## VI. EXPERIMENTS AND RESULTS

### A. Comparing stratagies with up-to-date information.

First of all we want to decide which strategy is better if we have up-to-date information every time the dispatcher makes a decision. In the Section IV Part C we showed that the Shortest Queue load balancing strategy is optimal for the system we use. The experiments results are consistent with theoretical result. The Shortest Queue strategy works much better than Random and Round Robin strategies. The results of the experiments are shown on the Figure 1.

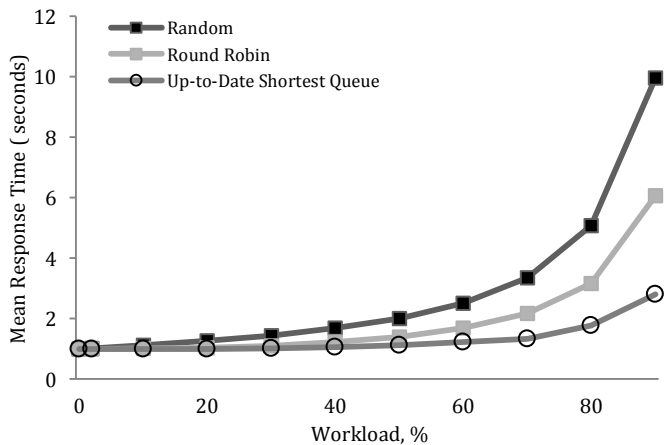

Figure 1. Increase of the mean response time depending on the system's workload for Random, Round Robin and USQ strategies.

### B. Using stale information in the dispatcher to make a decision.

It is practically impossible to get fresh information about the system every time we need to transfer a customer to a server. Thus, stale information is used in the real systems. Using of such information will affect badly the mean response time. The dependence of the mean delay from the length of the stale period (the time between information updates) is shown on the Figure 2. We can see that result become worse if the stale period is increased. The delay grows more significantly if the system is highly loaded. In this way the SSQ balancing strategy is similar to the strategies in which up-to-date information is used.

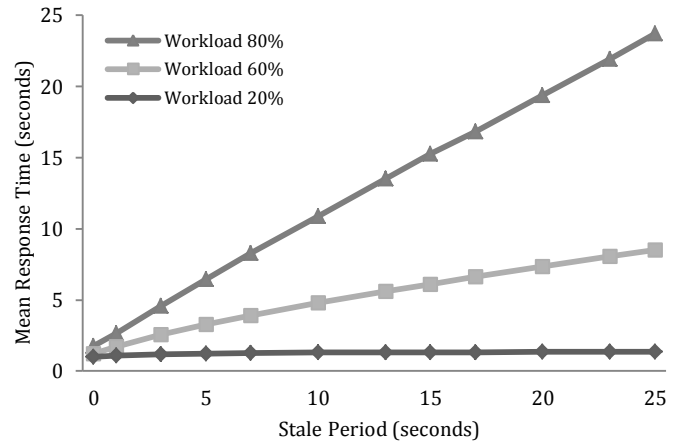

Figure 2. Mean response time for the SSQ balancing strategy depending on the stale period (the time between information updates). Different workloads of the system are presented.

Comparing the mean response time of the SSQ strategy to that of the Random strategy, we can see that it is worse even if the stale period is only 4 seconds. But with 2 seconds time between updates, SSQ strategy works better than Random, but still much worse than USQ which is considered as an ideal strategy. Figure 3 shows the results of the experiments that prove these statements.

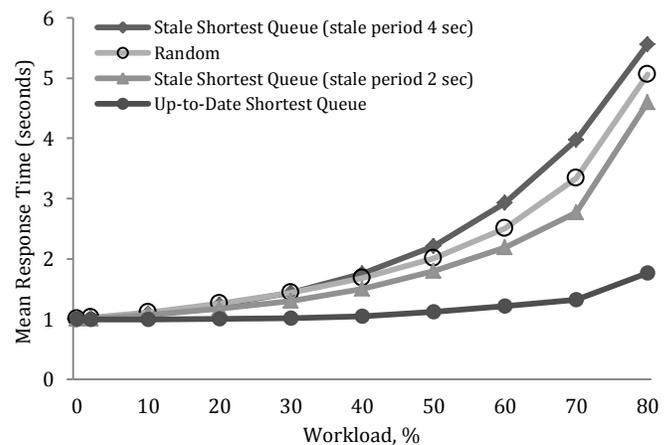

Figure 3. Comparison of the mean response time of the Shortest Queue balancing strategy with stale information to Random and Up-to-Date Shortest Queue strategies. Two different stale periods are shown.

### C. Studing the HSQ strategy performance.

The purpose of developing a new load balancing strategy is to decrease the mean response time of the system when stale information is used. Our experiments show that we have achieved this goal. Results, shown on the Figure 4, indicate that HSQ works much better than SSQ with the same stale period. Also we can see that the delay increases much slower when using the HSQ strategy instead of the SSQ.

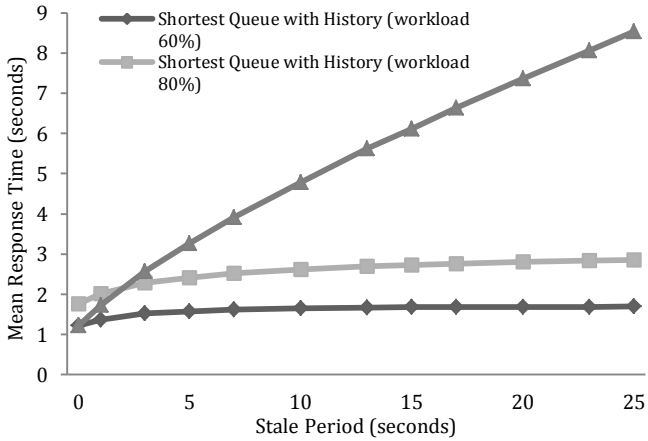

Figure 4. Mean response time for the Shortest Queue balancing strategy with stale information and history depending on the stale period (the time between information updates). Different workloads of the system are presented. The results are compared to the regular Shortest Queue with stale information.

Comparing the HSQ strategy to the USQ strategies, we can see that although it works worse than the optimal strategy the difference is not as large as for other strategies. The results of this comparison are shown on the Figure 5. We also want to mention that even with extremely long time between the updates the HSQ strategy performance is very close to those of the Round Robin strategy. It can be explained by careful examination of the strategy. If the stale period is very long and significant number of jobs arrives during it, then the HSQ strategy will work as follow:

*1)* Equalizes the lengths of the servers queues. As it transfers the job to the server with the shortest queue and then increments the length of the queue, it will give jobs to the servers with a less queue lengths until all the queue lengths become equal.

*2)* The strategy starts to work in round robin fashion since it has equlized the lengths of the servers queues. First we will randomly choose one server from 5, then one from the remaining 4, etc. Thus, we will schedule five next jobs to the five next servers and then will repeat this procedure from the beginning.

Thus, we can say that the value of the mean response time for the HSQ strategy in the case of a very long stale period will be very close to those of the Round Robin strategy.

VII. CONCLUSION

In this paper, we studied four load balancing algorithms using a simulation model consisting of one load balancer and five servers. Results show that shortest queue with up-to-date server load information gives the lowest delay as expected. But if the load/state information is significantly out of date then the shortest queue strategy performs worse even than random load balancing. To improve stale information based shortest queue, we have implemented a new history based shortest queue load balancing strategy. Although our new approach works worse than the optimal strategy, the difference is not as large as for other strategies. Even the state information is too old, our approach performs close to round-robin strategy. The improved strategy can be used in real system as it has low memory requirement and we do not have to update the state information too frequently to get reduced delay compare to other load balancing strategy discussed in the paper.

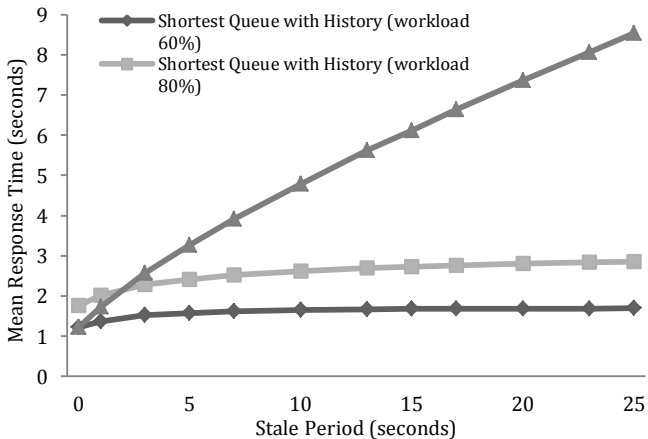

Figure 5. Comparison of the mean response time of the Shortest Queue balancing strategy with stale information and history to Round Robin and Up-to-Date Shortest Queue strategies. Two different stale periods are shown.